\documentclass[aps,12pt,tightenlines,amsmath,amssymb,showpacs,showkeys]{revtex4}

\def\Eqref#1{Eq.~(\ref{#1})}
\def\parenref#1{(\ref{#1})}
\newcommand{\be}{\begin{equation}}\newcommand{\ee}{\end{equation}}
\newcommand{\bea}{\begin{eqnarray}}\newcommand{\eea}{\end{eqnarray}}
\def\littlespace{$\;$}

\usepackage{graphicx}
\usepackage{bm}

\def\complexnumbers{\mathbb C}\def\realnumbers{\mathbb R}

\newcounter{remark}

\def\up{\hbox{\capseight UP}}
\def\down{\hbox{\capseight DOWN}}

\def\R{{\bm R}}
\def\i{{\bm i}}
\def\j{{\bm j}}
\def\k{{\bm k}}
\def\B{{\bm B}}
\def\r{{\bm r}}

\begin{document}

\title{\vskip -40pt\hfill$\hbox{\small
\rm Entropy \textbf{14}, \rm 665-686 (2012)}$\vskip10pt~\\
Experimental test of the ``special state'' theory of quantum measurement}

\keywords{special states, tests of quantum mechanics, retrocausality, Cauchy distribution}
\pacs{03.65.Ta, 05.40.Fb, 05.90.+m, 42.50.Xa} 

\author{L. S. Schulman}
\affiliation{Physics Department, Clarkson University, Potsdam, New York 13699-5820, USA}
\email{schulman@clarkson.edu}

\begin{abstract}
An experimental test of the ``special state'' theory of quantum measurement is proposed. It should be feasible with present-day laboratory equipment and involves a slightly elaborated Stern-Gerlach setup.

The ``special state'' theory is conservative with respect to quantum mechanics, but radical with respect to statistical mechanics, in particular regarding the arrow of time. In this article background material is given on both quantum measurement and statistical mechanics aspects. For example, it is shown that future boundary conditions would not contradict experience, indicating that the fundamental equal-a-priori-probability assumption at the foundations of statistical mechanics is far too strong (since future conditioning reduces the class of allowed states).

The test is based on a feature of this theory that was found necessary in order to recover standard (Born) probabilities in quantum measurements. Specifically, certain systems should have ``noise'' whose amplitude follows the long-tailed Cauchy distribution. This distribution is marked by the occasional occurrence of extremely large signals as well as a non-self-averaging property. The proposed test is a variant of the Stern-Gerlach experiment in which protocols are devised, some of which will require the presence of this noise, some of which will not. The likely observational schemes would involve the distinction between detection and non-detection of that ``noise.''  The signal to be detected (or not) would be either single photons in the visible and UV range or electric fields (and related excitations) in the neighborhood of the ends of the magnets.
\end{abstract}

\date{\today}

\maketitle

\section{Introduction\label{sintro}}

This article has three components. The first two are background for the third, which proposes---in some level of detail---an experimental test for the ideas propounded in the earlier sections. The components are
\begin{itemize}
\itemsep=-4pt 

\item A theory of quantum measurement that is conservative: there is \textit{only} unitary time evolution. There is no wave function collapse, there is no need for ``many world'' concepts, and the wave function is \textit{not} merely a construct for calculating probabilities.

\item A modification of statistical mechanics that is radical but which contradicts no experience or experiment. Because of future conditioning, many initial conditions are excluded, contrary to standard statistical mechanics. In particular there is a form of conditioning that can be used to motivate the quantum ideas. A rationale for this future conditioning introduces cosmological considerations.

\item A modified Stern-Gerlach experiment in which physical phenomena not predicted by the Copenhagen interpretation would occur. In particular the test would be conducted using two protocols, in one of which there would be an observable signal, in the other there would not. That signal could be the emission of photons in the eV range or the appearance of electric fields (and related effects) near the ends of the magnets.

\end{itemize}

With respect to the background material, there will be frequent reference to \cite{timebook} and indirectly to the many citations therein. After publishing \cite{timebook}, I did not much work on this problem. Although there are many open theoretical issues, I felt that there could only be progress if these ideas were tested experimentally. If they passed, there would be no shortage of thought devoted to them.

What has rejuvenated my interest is the possibility of an experiment that could confirm features of this theory. The trail to the practical suggestion contained in the present article began with the implementation in \cite{jacques} of the Wheeler delayed-choice experiment \cite{wheeler, wheeler1981}. Although Wheeler made his prediction entirely within the framework of the Copenhagen interpretation, there is an apparent inversion of causality that suggested the kind of tightly interconnected future and past that characterizes my own work. In conference presentations in which I reported preliminary ideas on this subject \cite{sandiegoretro, note:CTS} I focused on that experiment; but I later realized that the essential physical feature that would allow the experimental test did not depend on the ``delayed'' part of the story. This makes the experiment much easier.

The sections of this article follow the enumeration given above, quantum mechanics (Sec.~\ref{squantum}), statistical mechanics (Sec.~\ref{sstatistical}), experiment (Sec.~\ref{sexperiment}). Sec.~\ref{sdiscussion} is a discussion.

\section{Quantum Mechanics\label{squantum}}

This is a brief and selective summary of my quantum measurement ideas, based on the central notion of ``special states'' (henceforth mostly sans quotation marks).

Consider the Schr\"odinger cat. For this unfortunate feline, if the trigger of the device aimed at it depended on, say, an atomic decay, the probability of a living cat would be the non-decay probability, say 1/2, for the time interval set for the ``experiment.'' I will now give an example where---if you could prepare the microscopic state of the apparatus---you could keep the cat alive.

\begin{figure}
\centerline{\includegraphics[height=.3\textheight, width=0.8\textwidth]{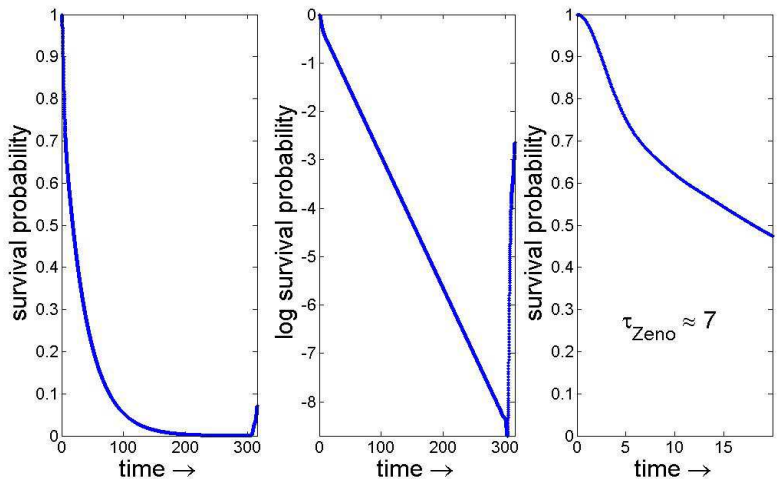}}
\caption{Normal decay. ``$N$'' (the size of $\Omega$ in \Eqref{edecayhamiltonian}) is 100, and at about time-300 there is a Poincar\`e recurrence due to this finite dimension. The semilog plot shows excellent exponential decay up until then. On the right is early-time non-exponential decay (note the shorter times plotted), related to the so-called quantum Zeno effect. The calculated ``Zeno time,'' 
$\tau_{\hbox{$_{\!_\mathrm{Zeno}}$
}}\equiv
\frac
\hbar{\sqrt{\langle \psi|H^2|\psi\rangle-\langle\psi|H|\psi\rangle^2}}
$, is about 7.}
\label{foneleveldecay}\end{figure}

First consider the formalism for ordinary decay. A single level decays, emitting a photon. For a finite-time context there will be a band of energies into which it can decay, and this is modeled as a finite number, $N$, of narrowly spaced levels (so $N\gg1$). A Hamiltonian for this system is
\be
H=
 \left(
\begin{array} {cc}
  \omega          & \phi \\
  \phi^\dag  & \Omega
\end{array}
\right)
\,,
\label{edecayhamiltonian}
\ee
where $\omega\in\realnumbers$, $\phi\in\complexnumbers^N$, $N\gg1$, and $\Omega$ is a real, diagonal $N\!\times\! N$ matrix. The wave function, $\psi$, is an $(N+1)$-row column vector and initially has 1 in its first entry, zeros elsewhere. The survival probability is $S(t)\equiv |\langle\psi(0)|\exp(-iHt/\hbar)|\psi(0)\rangle|^2$\@. A numerical calculation of this quantity provides the graphs of Fig.~\ref{foneleveldecay}. The semilog plot shows that the decay is close to exponential until $t\approx300$, at which point a (quantum) Poincar\'e recurrence sets in, due to the finiteness of $N$ (100 for this calculation). I also show the early-time quantum Zeno effect, manifested as initial non-exponential decay. The time interval during which this is significant matches well to the ``Zeno time'' that I proposed in
\cite{jumpduration, contin, jumppassage2ndedition}.

We next suppose that the decaying atom is one of many, all of which have essentially the same matrix elements for decay with photon emission. The number atoms (and the number of associated levels) is $n$, and we assume $N\gg n\gg1$\@. This more general Hamiltonian is again given by \Eqref{edecayhamiltonian}, but the meaning of the symbols has changed. Now $\omega$ is an $n\!\times\!n$ matrix, constant (all the atoms are the same), and diagonal. The coupling, $\phi$, is now a rectangular $n\!\times\!N$ complex (in general) matrix, while $\Omega$ is as before.

\begin{figure}
  \includegraphics[height=.3\textheight, width=.6\textwidth]{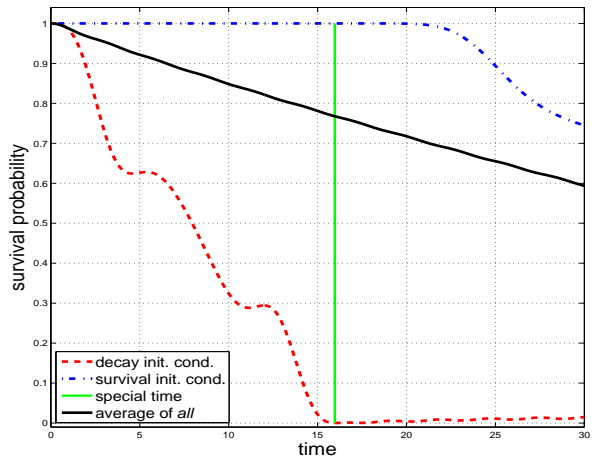}
  \caption{Decay from a collection of $n$ similar levels. The non-solid lines show the special states, which take values close to 0 and 1 at the selected time,~16.}
\label{fmultileveldecay}\end{figure}

The atoms are assumed close enough and steady enough to interact coherently and their net excitation number is one; hence the wave function has $N+n$ components, and the initial condition (non-decay) requires that all non-zero elements of the initial wave function lie in the first $n$ entries. The resulting decay \cite{linear} is remarkable and is shown in Fig.\ \ref{fmultileveldecay}. The \textit{average} decay is shown in the solid (essentially) straight line (black in color). This is relatively normal, although the linearity (as opposed to exponential decay) is due to particular circumstances. But what are not normal are the blue (dashed) and red (dash-dot) curves.

Those curves require explanation. First a time, $t_0$, is picked. In this case, it's 16 (shown as a vertical green line in the figure). Then, by a method described in App.~\ref{smethod}, I find two special classes of states: those for which $S(t_0)\approx0$ and those for which $S(t_0)\approx1$\@. That appendix tells you how to find the \textit{initial} conditions, the $\psi(0)$'s, that lead to the all-decayed or all-not-decayed states at time-$t_0$\@. (That there are such states is equivalent to my demand for special states, as we shall see.) The blue (dash-dot) curve in the figure is the time-dependence of an all-not-decayed-at-$t_0$ state. At $t_0$ it is essentially still in the initial subspace of undecayed states. The red (dash) curve is the time dependence of one of the other class of states, those that are nearly fully decayed at time-$t_0$\@. For both, after $t_0$ there is no 0-1 requirement, although by continuity they do not change radically. (Regarding imperfect attainment of 0 or 1, see~\cite{timebook}.)

\subsection{Use of the special state\label{suse}}

We return to the Schr\"odinger cat. Suppose the gun fires in response to a system of atoms of the sort described above. Let the full wave function at time zero be $\Psi$; this includes the cat, the cage, the weapon, the atom that triggers the weapon---everything! Let the Hamiltonian for all this be $H_\mathrm{total}$ and let the time at which we ``look to see if the cat's alive'' be $t_0$\@. In some reasonable approximation, the wave function can be written $\Psi=\Theta A\Phi$ with $\Phi$ the cat wave function, $A$ the wave function of the atoms (and their decay products) and $\Theta$ referring to everything else. We suppose that the time $t_0$ is such that there is a 50-50 chance that the cat is alive. Letting $U(t_0)\equiv\exp(-iH_\mathrm{total}t_0/\hbar)$, this situation can be schematically written
\be
\Psi(t_0)=U(t_0)\Psi(0)=\alpha\Theta_\ell(t_0) A_\ell(t_0)\Phi_{\ell}(t_0)+\beta\Theta_d(t_0) A_d(t_0)\Phi_{d}(t_0)
\label{ecat}
\,,
\ee
($\ell$ and $d$ are ``living'' and `dead') with $|\alpha|^2\approx|\beta|^2\approx1/2$ (the $\Theta$'s $A$'s and $\Phi$'s are normalized to 1). This state is what Griffiths \cite{griffiths} has called ``grotesque,'' a superposition of macroscopically different states. Give the decay example just discussed, it is clear how to keep the cat---definitely---alive. Start the atom wave function in one of the blue, non-decaying, states shown in Fig.\ \ref{fmultileveldecay}, call it $A'(0)$\@. This means that there is a state of the whole thing, call it $\Psi'(0)$, such that
\be
\Psi(t_0)=U(t_0)\Psi'(0)=U\left\{\Theta(0) A'(0)\Phi(0)\right\} =\Theta_{\ell}(t_0) A_\ell(t_0) \Phi_{\ell}(t_0)
\,.
\label{ecatliving}
\ee
Similarly, there are red---full decay---states (call them $A''$) with corresponding $\Psi''$ such that
\be
\Psi(t_0)=U(t_0)\Psi''(0)=U\left\{\Theta(0) A''(0)\Phi(0)\right\} =\Theta_d(t_0) A_d(t_0) \Phi_{d}(t_0)
\,.
\label{ecatdead}
\ee
We have thus obtained definite, non-grotesque, states without any black magic of ``measurement;'' the only thing that happens is pure, unitary time evolution.

\subsection{The assumption concerning special states in Nature\label{sassumptionspecial}}

The major assumption concerning special states is that in every situation in which there might emerge grotesque states (and this goes beyond human laboratory experiments) the initial conditions are special.

This assumption implicitly makes two claims. The first is less radical: there are enough special states to do the job. Within any apparatus capable of creating grotesqueness there are enough degrees of freedom for certain rare states to give definite (non-grotesque) outcomes. I cannot verify this in general but I have explored many models of apparatus and found special states for all of them. Interestingly, circa 1990---after 60 years of debate---there wasn't a single apparatus model that I could find \cite{note:bona} that was realistic enough to be used to address this question. It was for this reason that Bernard Gaveau and I developed a quantum apparatus model \cite{model} in which I could subsequently find special states.

The other claim is truly radical. It places a restriction on \textit{initial} conditions. The fundamental axiom of statistical mechanics states that, given a m\textbf{\large a}croscopic description of a system, the m\textbf{\large i}$\mskip1mu$croscopic states associated with it are \textit{all} those consistent with the macroscopic description.

I say, no, you don't take \textit{all} states, you only select very particular ones, those that are what I call special, namely those that do not lead to grotesqueness.

To provide perspective on this claim I will discuss the arrow of time, since connected to that notion there is also a tremendous elimination of microscopic states. From that discussion will emerge a context in which justification of my selection claim can be imagined.

However, this does not exhaust the tasks imposed for the recovery of the standard results of quantum measurements. In particular, the existence of special states does not by itself give the Born probabilities. It is this requirement---on which we next focus---that leads to the experimental test proposed in this article.

\subsection{Recovering probabilities\label{sprobabilities}}

In classical mechanics if you know an initial phase space point, the outcome at any later time is certain. You use probability when there are many initial points consistent with the information you have (so probabilities other than unity mean your information is incomplete) \cite{note:calculation}. In this case the probability of any particular outcome is proportional to the volume of phase space that leads to that outcome. This can be considered a corollary of the arrow of time definition given earlier. (For working backward---retrodiction---Bayesian rules enter, but that's another story \cite{implication}.) The quantum version of this replaces volumes of phase space by the dimensions of subspaces of Hilbert space and also---it is said---introduces another kind of probability, one that supposedly is intrinsic.

For the ideas expressed earlier using special states there is no additional layer of probability. The collection of special states for any particular outcome forms a vector space, and my postulate is that the probability of a given outcome is proportional to the dimension of the associated vector space of special states. This is a bold postulate, since the usual Born probabilities depend little on the apparatus and are computed from the wave function of the system being measured. On the other hand, I require identification of the special states of system and apparatus combined and a counting of (vector space dimension of) those special states. I have not managed to check this even in some of the apparatus models where I've succeeded in finding special states. I believe the reason is that the special states I've solved for are atypical; after all, one should not expect solvability to be an attribute of a real measurement apparatus.

However, the idea that I could exhibit and count special states is an optimistic one. Some years ago I took the opposite view and in a fit of pessimism said, suppose I could have any special states I wanted, what constraints on their distribution would I have, and---maybe---those constraints would mean the whole idea was wrong.

In the following discussion I'll make the assumption that Nature, the environment, the apparatus, even parts of the system being measured, can provide the rare microstates needed. Moreover, there will be many for each outcome~\cite{note:manyrare}. To see how this assumption is implemented and the constraints it imposes, it will be useful to focus on a particular experiment.

\def\up{{\small\textsc{up}}}
\def\down{{\small\textsc{down}}}

Consider a Stern-Gerlach experiment measuring the $z$-component of an atom's spin. For an atom having net spin 1/2, let the prepared wave function be
\be
u_\theta = e^{i\theta\sigma_x/2} {1\choose 0}
\,,
\label{einitialsterngerlach}
\ee
with $\sigma_x$ the Pauli spin matrix. Only two outcomes are possible, designated \down\ and \up. Their detection involves a hot wire detector downstream from the magnet supplying the inhomogeneous field that induces the measurement (coupling spin and translational degrees of freedom). The standard prediction is that they come with the ratio
\be
\tan^2 {\theta\over2}={\sin^2\theta/2\over\cos^2\theta/2}
\,.
\ee
Now a special state that will send the atom to the right place for, say, a \down\ measurement, may be rare, but we need to look for the \textit{least} rare among all those that could do the job. These least-rare states will presumably be unusual states of the environment, but let's consider where, spatially, that rarity will be manifested. After the atom has passed through the magnet it would be necessary to coherently recombine the spatially separate portions of the wave function, while if a rare environmental state were available prior to the atom's deflection by the magnetic field it would only have to rotate the spin by, say, $\frac\pi2-\theta$\@. So I'll assume that the least unlikely states are those that act on the spin wave function in the following way:
\be
u_\theta = e^{i\theta\sigma_x/2} {1\choose 0}\to
e^{i\psi\sigma_x/2}e^{i\theta\sigma_x/2} {1\choose 0}
=e^{i(\psi+\theta)\sigma_x/2} {1\choose 0}
\label{ekick}
\,.
\ee
I'm about to make a slight shift in perspective. Instead of counting actual microstates of the environment I'll sort them by their effect, namely by the size, $\psi$, of the rotation they can induce on the wave function. Suppose that there are $f(\psi)$ states that can rotate by angle $\psi$\@. Without loss of generality for what follows we can normalize the function $f$ so its integral over $\psi$ is 1\@. In any given experiment the net result of all these special states will be a rotation by $\sum_{\alpha=1}^N\psi_\alpha$ if the spin is subject to $N$ such rotations/kicks/special states along its path.

Dealing with this observation almost led me to abandon my quantum measurement ideas. If you imagine that a large number of ``kicks'' (rotations of the sort discussed) are necessary then one would expect to be able to use the central limit theorem, in which case the relative ratios for getting \up\ or \down\ would be a ratio of Gaussians, \textit{not} the tangent-squared function given earlier.

It turns out that this problem has a solution and its solution is a key to the experimental test that I here propose. First, drop the assumption that you can use the central limit theorem, i.e., we do \textit{not} assume that the function $f$ has a second moment. This leads us into the world of the L\'evy distributions, with many peculiar properties, as we shall see. Let us assume that whatever happens to our spin must happen in a single ``kick.'' It follows that the function $f$ must satisfy
\be
\tan^2 {\theta\over2}
              =  {F(\theta + \pi)\over F(\theta)}\,,
\ee
where
$F(\theta) \equiv \sum_{k=-\infty}^{\infty} f(\theta + 2k\pi) $. (Note that $F(\theta)$ gives all ways of getting $u_\theta$ to become \up, and $F(\pi-\theta)$ gives the ways to become \down. The solution to this functional equation is
\be
f(\psi)=C_a(\psi)\,,\quad \hbox{with~} C_{a}(x) = {a/\pi \over x^2 + a^2}
\ee
for $a$ small. So you can do it! (And, this distribution has the property that if the sum of $n$ samples drawn from it is far larger than  $na$, the least unlikely way to do this is a single large kick, all the others much smaller. For the Gaussian they'd all be about the same size.)

For further discussion of the function, $C_a$, the Cauchy distribution, see \cite{timebook}, as well as many books on probability theory, e.g., \cite{taqqu1, taqqu2}.

As to the parameter $a$, if it is too large, deviations from standard probabilities will be observed, but since I don't know where this noise is coming from I cannot use this for experimental predictions. I also mention that the demonstration above can be extended to many dimensional choices, not just spin-1/2 and not just spin. See~\cite{timebook}.

\section{Statistical Mechanics\label{sstatistical}}

\subsection{The arrow of time\label{sarrow}}

The usual statement of the thermodynamic arrow of time is that entropy increases, that is, it increases in one time direction, not the other. Alternative ways of saying this exist, for example the impossibility of converting heat to work. I will give another formulation, one that focuses on assumptions on microscopic states. How does one predict? If you isolate a glass of water containing ice cubes at 2 p.m., your prediction on its form at 3 p.m.\ is based on assuming that all microscopic states consistent with what you see at 2 p.m.\ are equally likely. In principle you evolve these forward in time and average, with the vast majority of microstates giving smaller ice cubes, colder water. If this system had been isolated since 1 p.m.\ your estimate of its 1 p.m.\ state would be based on an entirely different method. Using your 2 p.m.\ information, you make a guess about what it might have been at 1 p.m.\ and evolve that forward (as you did from 2 to 3 p.m.). If it fits what you see at 2 p.m., then it's a possible 1 p.m.\ state. (Had you propagated back all the 2 p.m.\ microstates, you would find smaller ice cubes, colder water at 1 p.m., which contradicts experience.) These different rules are an alternative statement of the arrow of time.

Now consider what you have implied about the 2 p.m.\ microstates. If you view them as initial conditions, everything goes, all of them are OK\@. But if you view them as having evolved from an earlier condition, macroscopically specified, then almost all of them are rejected. How do I know this? I can appeal to the usual formulation, the increase of entropy. The number of microstates is given by $\exp(S/k_B)$ with $S$ the entropy. Lower entropy at the earlier time means fewer states, and if I plug in plausible numbers for water and ice in a normal size glass, you will find that the rarity of the 2 p.m.\ microstates, considered as final states, is astounding, numbers like one in $10^{10^{24}}$\@.

\subsection{The cat map\label{scatmap}}

\begin{figure}
  \includegraphics[height=.25\textheight]{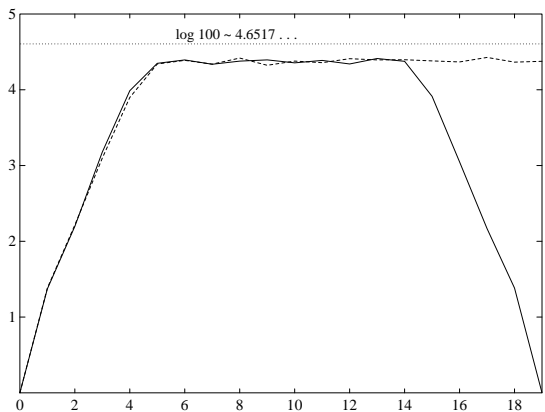}
  \caption{Entropy as a function of time for a cat-map simulation with a cryptic constraint at time 19 and for a simulation with no cryptic constraint. 100 coarse grains are used. Note that there is essentially no difference in the initial behavior.}
\label{fentropytwo}\end{figure}

Another example illustrates how a selection of \textit{initial} microstates can take place. Consider the ``cat map,'' an area-preserving transformation of the unit square that has served as a model of equilibration \cite{arnold}. The mapping is
\bea
x'&\equiv& x+y  \,, \ \hbox{mod}~1 \\
y'&\equiv& x+2y \,, \hbox{mod}~1
\,.
\eea
A collection of points (thought of as ideal gas particles) starting out in a small region of the square will rapidly spread throughout. The equilibration can be quantified by coarse graining the unit square and replacing a microscopic specification (giving the exact position of each point) by simply listing the number of points in each grain. The entropy is then defined as $S\equiv-\sum p_\alpha \log p_\alpha$ with $\alpha$ labeling the equal area (by construction) grains and $p_\alpha=n_\alpha/n$ with $n_\alpha$ the number of points in grain-$\alpha$ and $n$ the total number. The expansion of such a gas is illustrated in Fig.\ \ref{fexpansion1} and the associated entropy increase shown in Fig.\ \ref{fentropy}. Next I show the continuation of the time evolution. Fig.\ \ref{fexpansion2} shows the evolution of the same points for later times, and the associated entropy dependence appears in Fig.~\ref{fentropyall}
\def\hs{20}
\begin{figure}
\centerline{
  \includegraphics[height=.1\textheight]{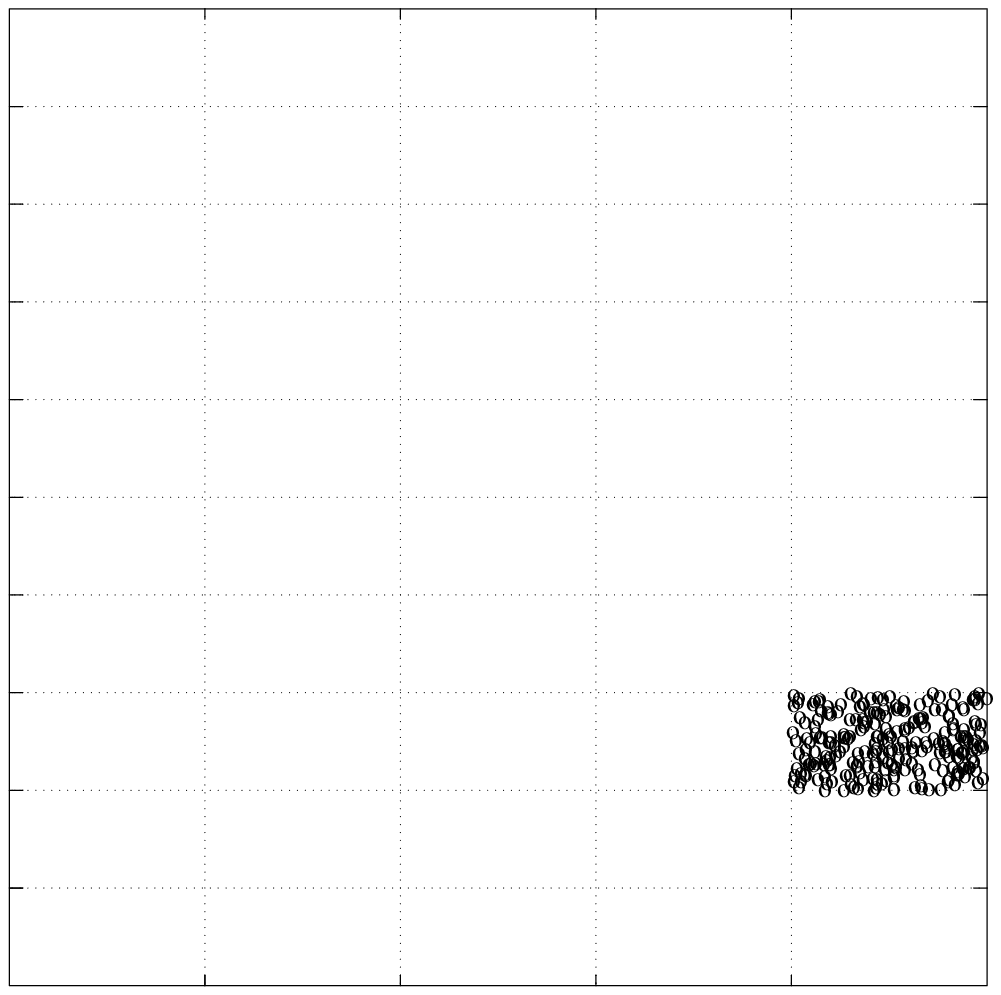}\hskip -\hs pt
  \includegraphics[height=.1\textheight]{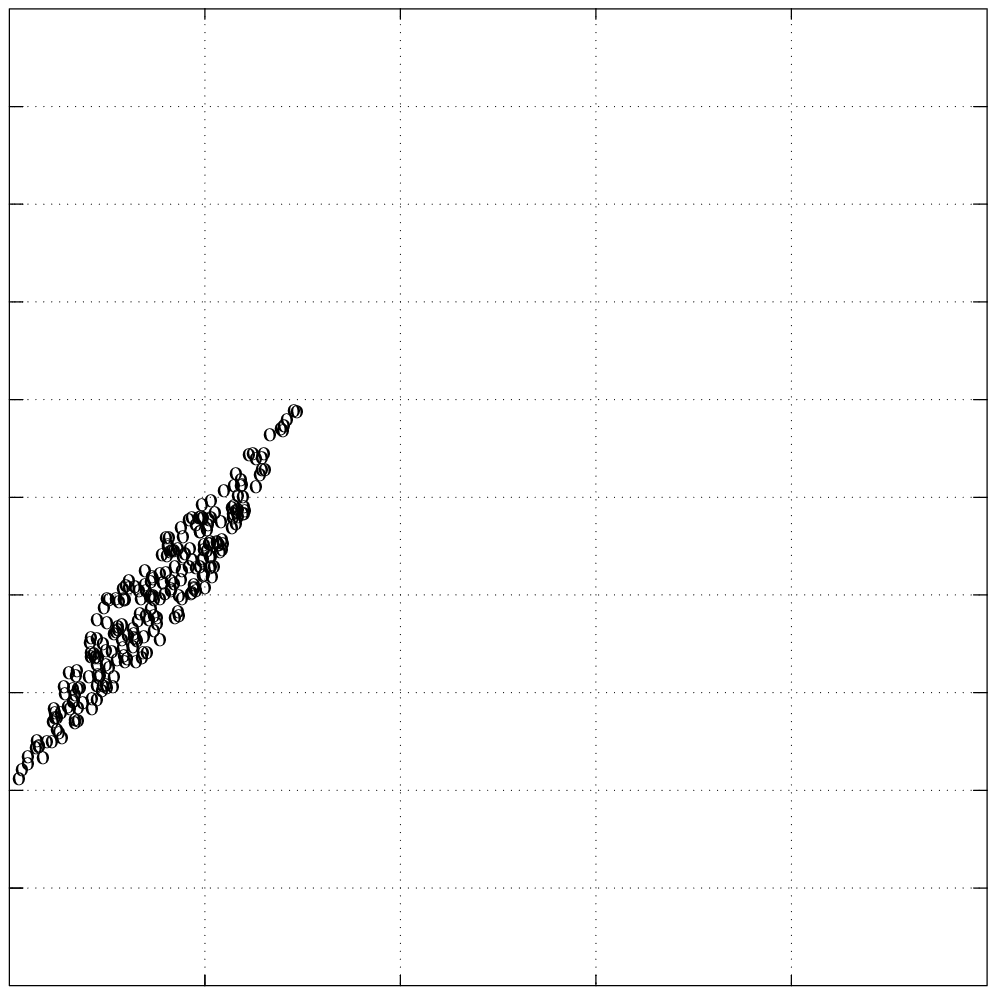}\hskip -\hs pt
  \includegraphics[height=.1\textheight]{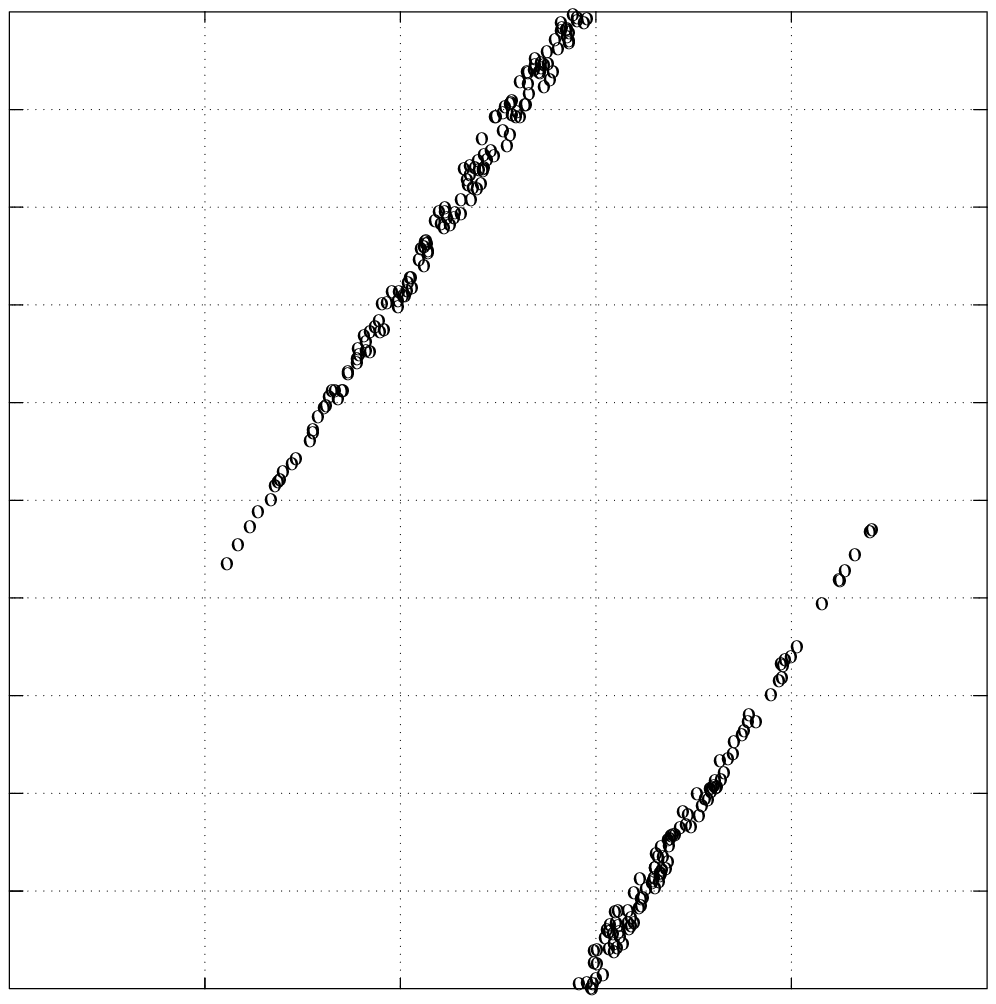}\hskip -\hs pt
  \includegraphics[height=.1\textheight]{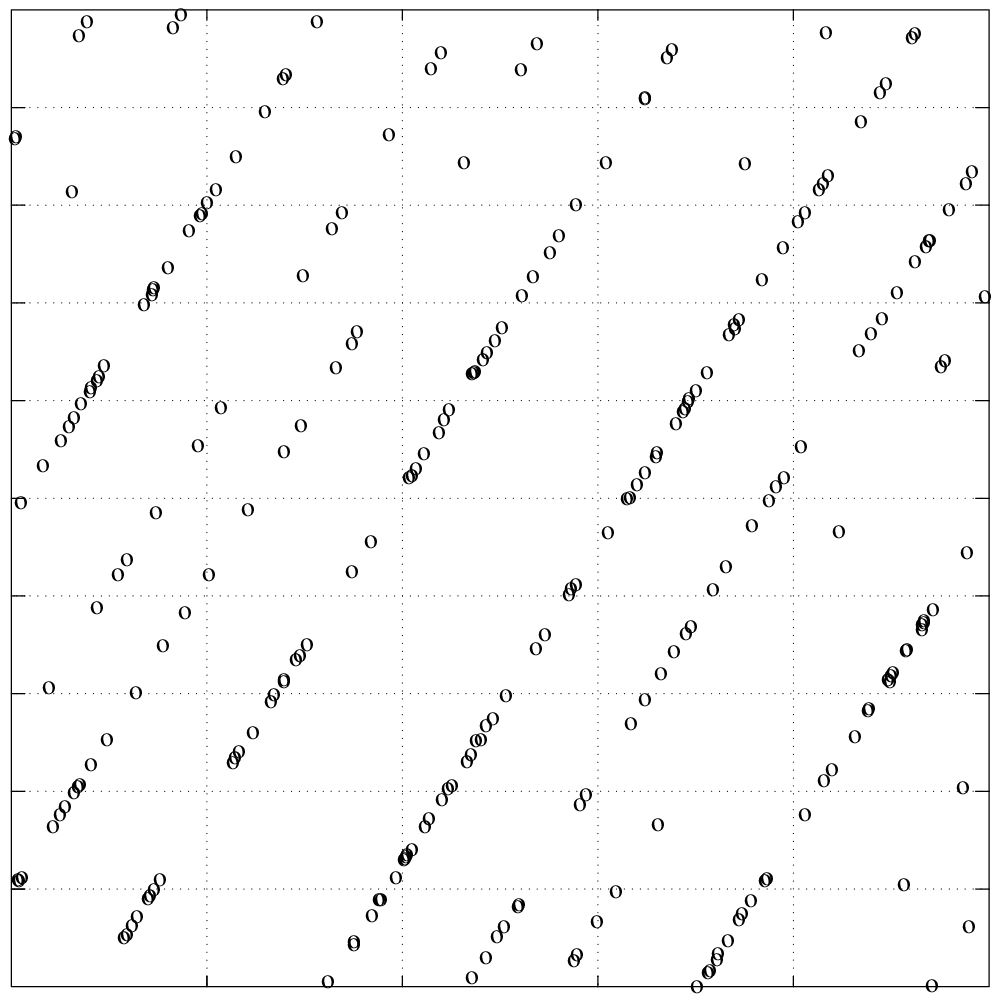}\hskip -\hs pt
  \includegraphics[height=.1\textheight]{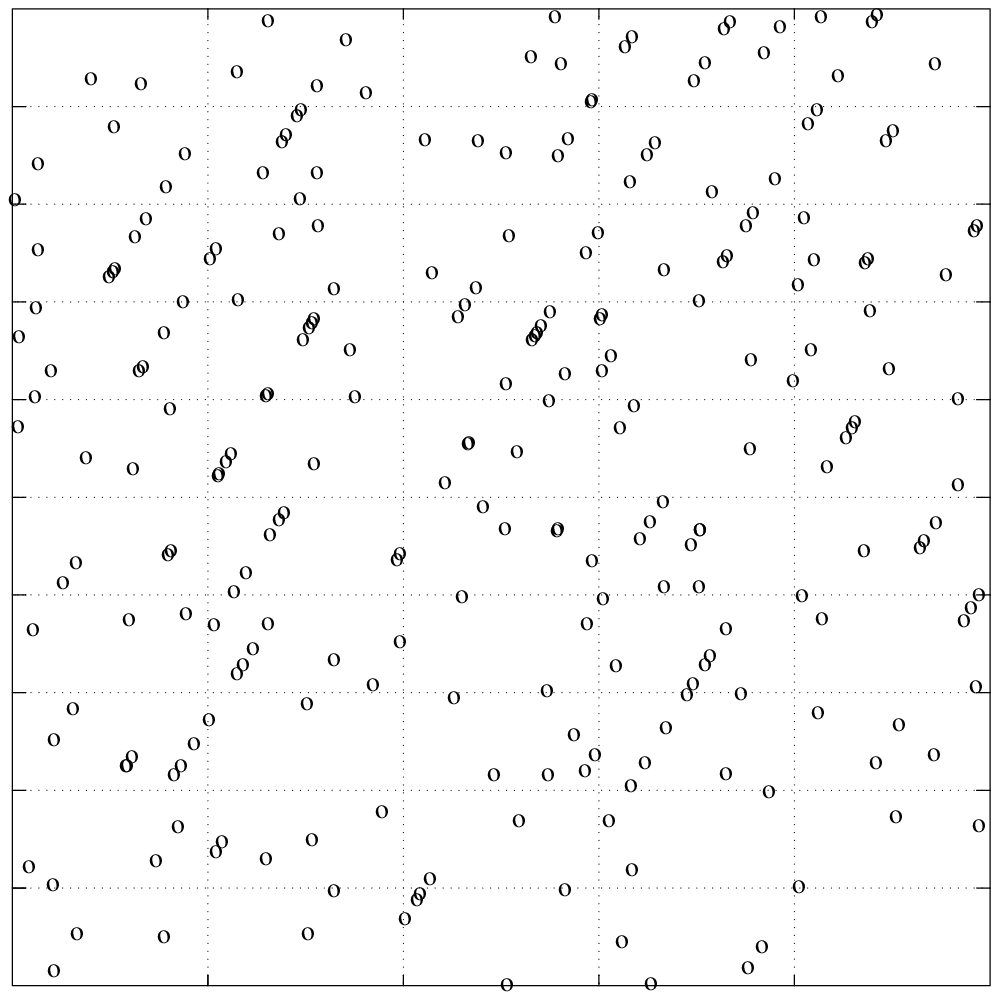}\hskip -\hs pt
  \includegraphics[height=.1\textheight]{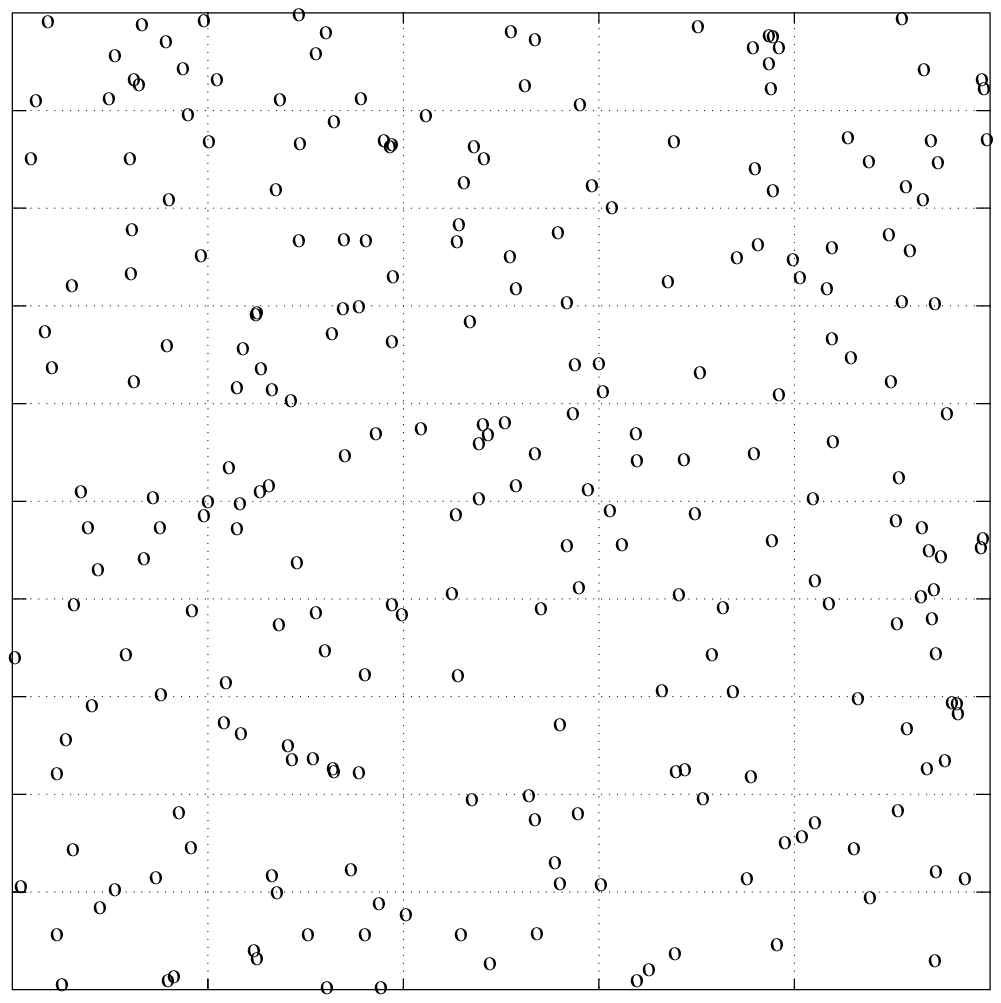}
}
  \caption{Times 0, 1, 2, 4, 5, 8 in the evolution of a gas of 250 particles under cat-map dynamics.}
\label{fexpansion1}\end{figure}
\begin{figure}
  \includegraphics[height=.25\textheight]{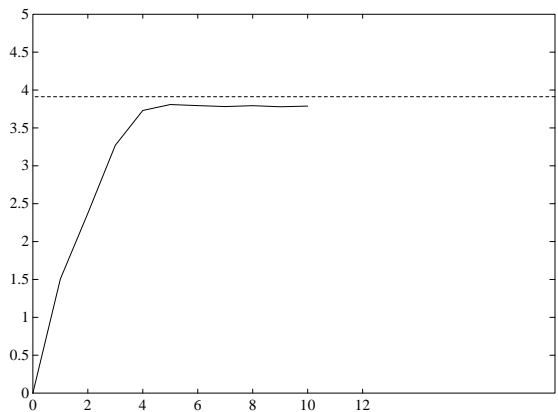}
  \caption{Entropy as a function of time for the expanding gas of Fig.~\ref{fexpansion1}. }
\label{fentropy}\end{figure}
\begin{figure}
\centerline{
  \includegraphics[height=.1\textheight]{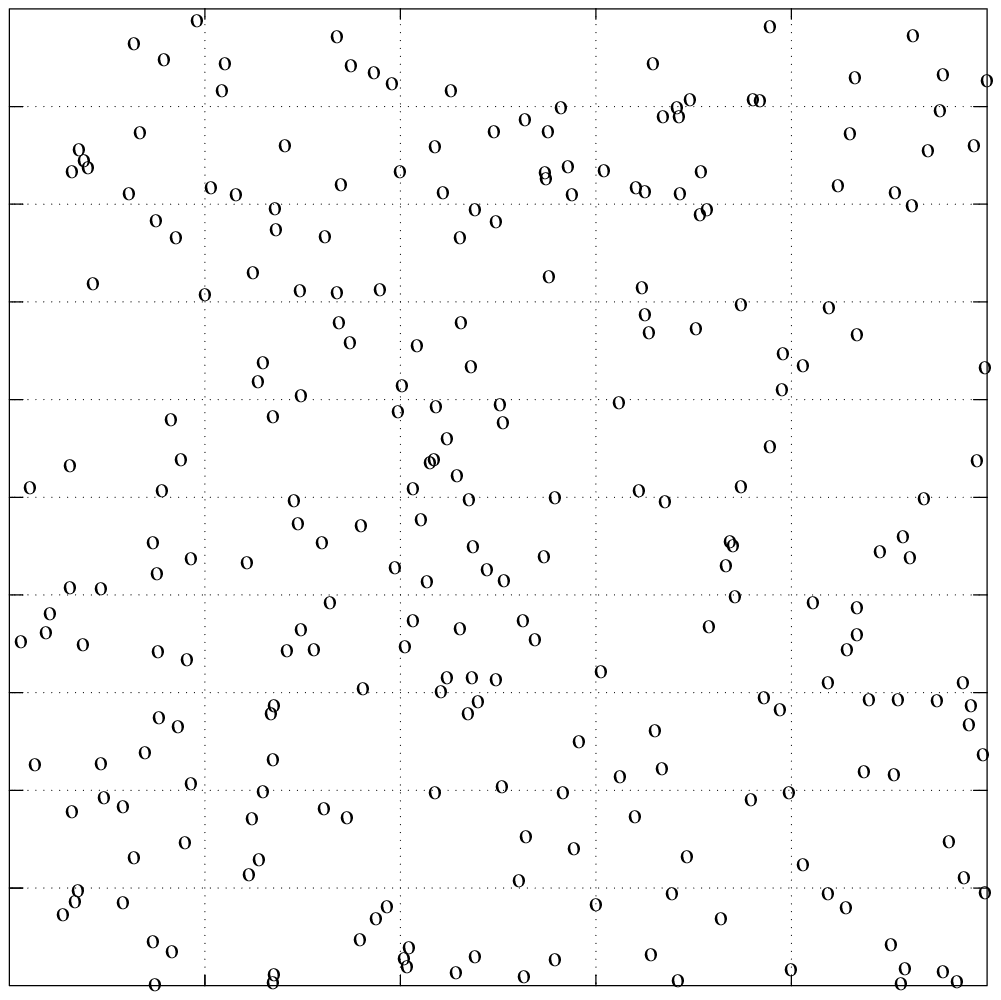}\hskip -\hs pt
  \includegraphics[height=.1\textheight]{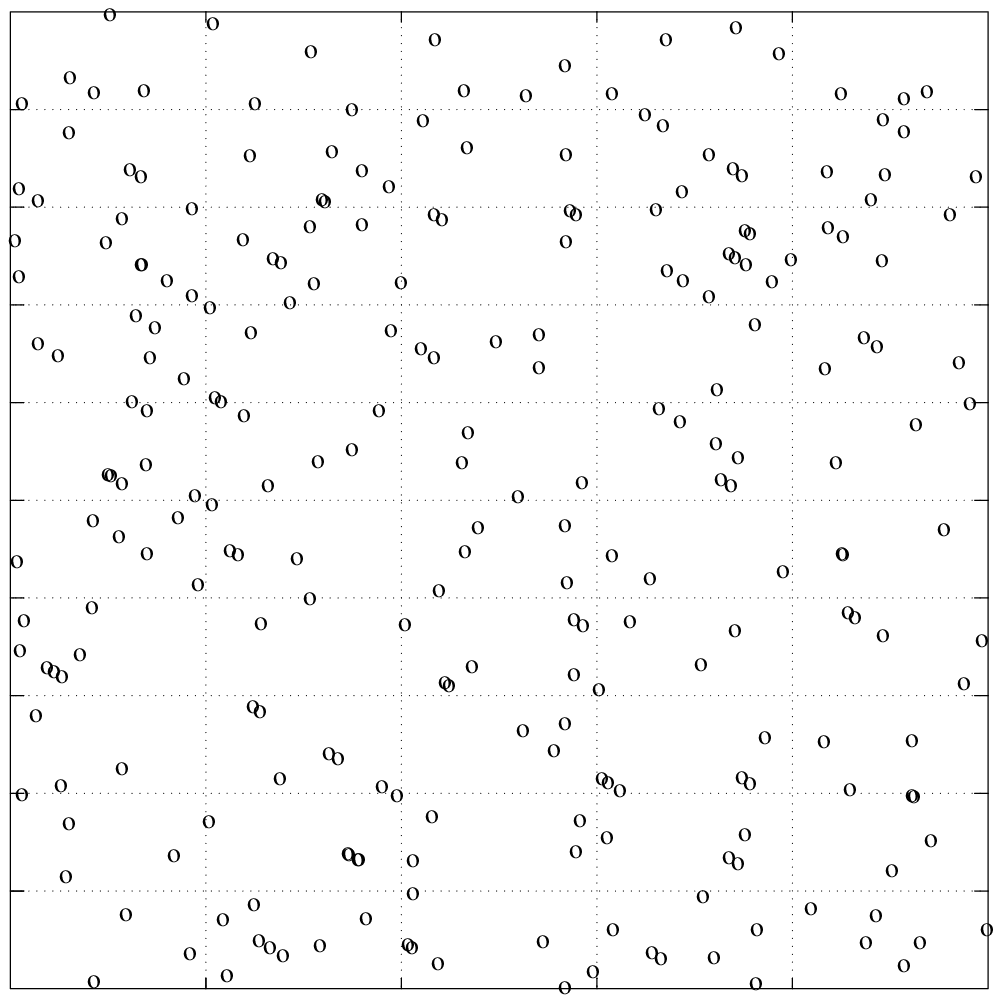}\hskip -\hs pt
  \includegraphics[height=.1\textheight]{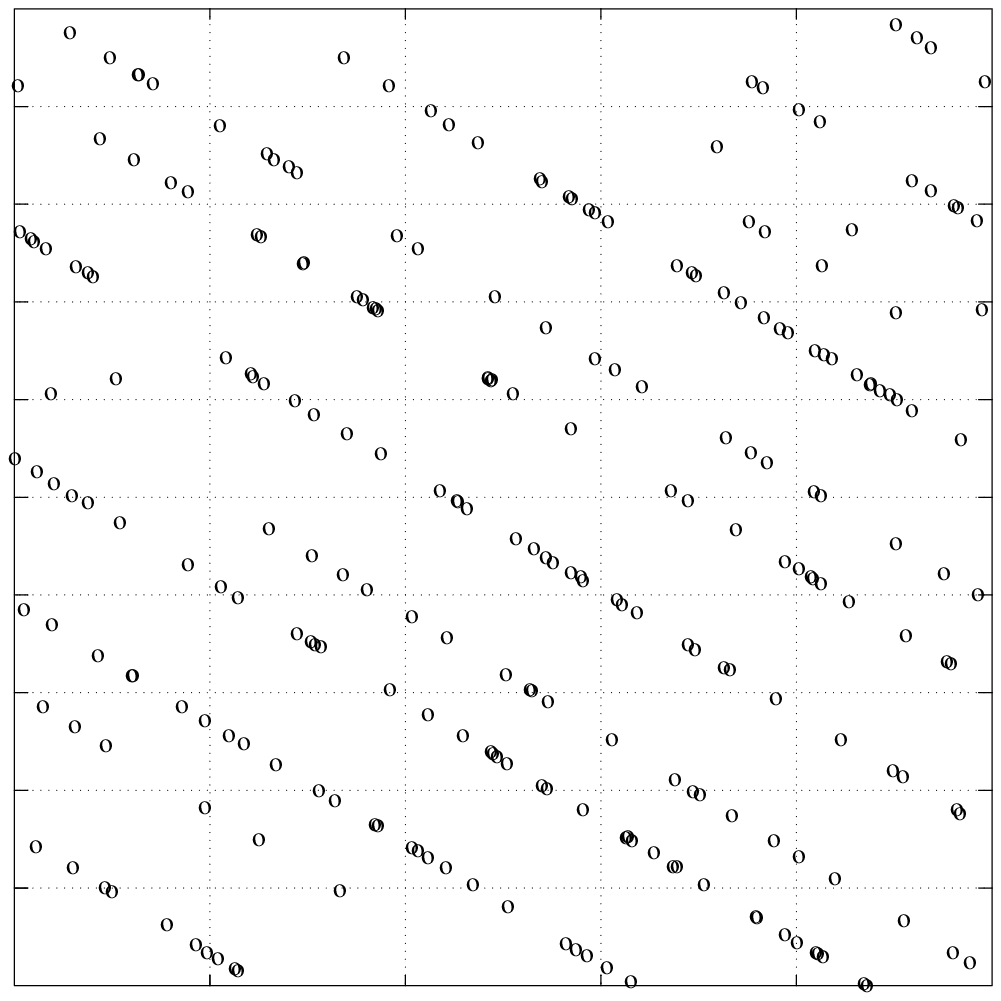}\hskip -\hs pt
  \includegraphics[height=.1\textheight]{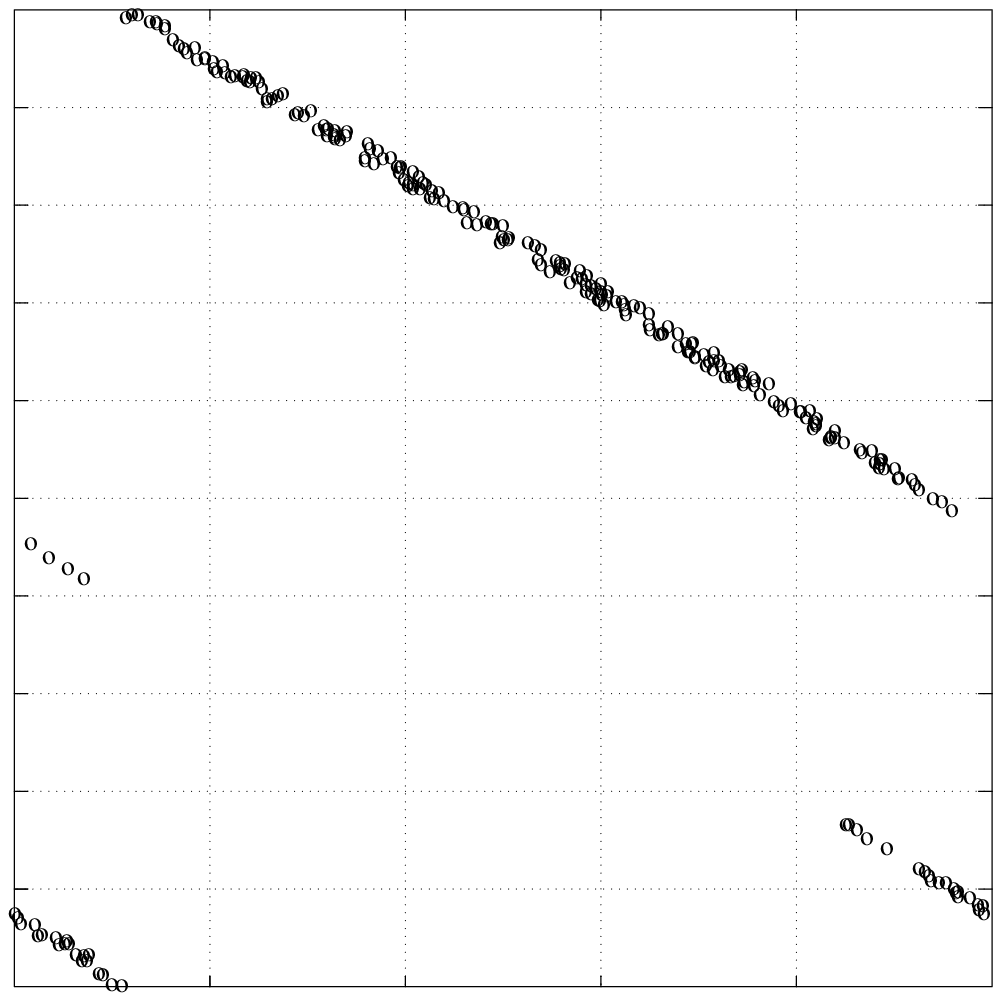}\hskip -\hs pt
  \includegraphics[height=.1\textheight]{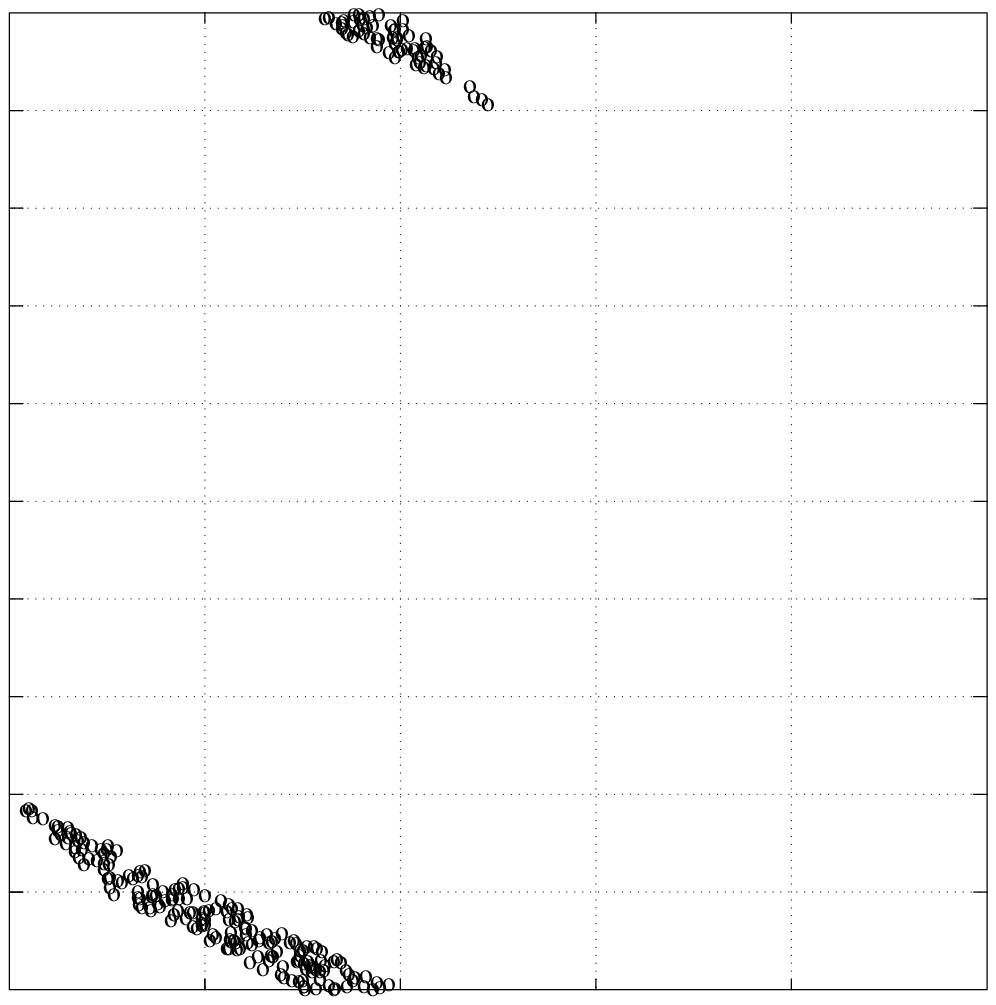}\hskip -\hs pt
  \includegraphics[height=.1\textheight]{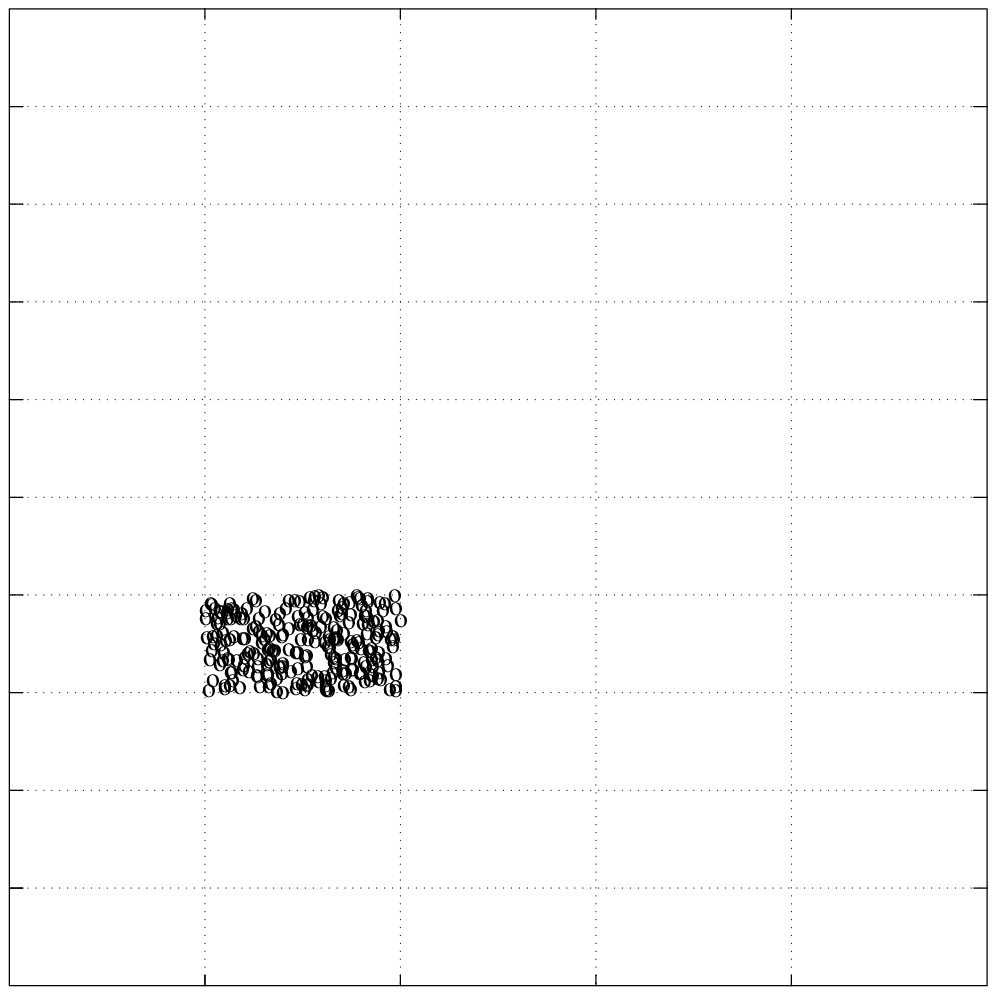}
}
  \caption{Times 11, 14, 15, 17, 18, 19 in the evolution of a gas of 250 particles under cat-map dynamics.}
\label{fexpansion2}\end{figure}
As you can see, something funny is going on. Instead of having the particle locations continue to fluctuate they come back together and the entropy decreases. This is not the result of a lucky Poincar\`e recurrence, which would only occur after the order of $50^{250}$ time steps. Rather, I solved a \textit{two}-time boundary value problem, finding points, all of which were gathered in a single box, at times separated by 19 time steps. This means that the point locations at time-0 were not at all random, even though they appear to be. They have a cryptic constraint. This constraint is mild by earlier standards, ruling out a mere 98\% of all points rather than $1-1/10^{10^{24}}$\@. Another important point, illustrated in Fig.\ \ref{fentropytwo}, is that the initial behavior of the macroscopic quantity, entropy, is the same with or without the cryptic constraint (that figure is for another simulation in which 100 coarse grains were used).
\begin{figure}
  \includegraphics[height=.25\textheight]{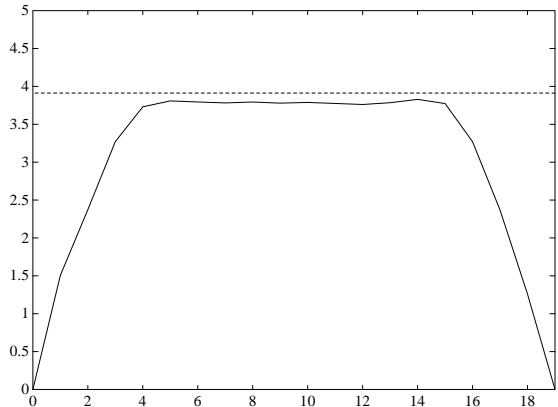}
  \caption{Continuation of entropy as a function of time, including the contracting segment, as shown in Fig.~\ref{fexpansion2}. }
\label{fentropyall}\end{figure}

The conclusions I draw from this example are as follows. With future boundary conditions you can restrict the set of initial conditions. Moreover, you can't tell the difference. The implication is that \textit{the usual axiom of statistical mechanics, equal probability for all microstates, is far stronger than is justified by experience.}

\subsection{Special states and determinism\label{sdeterminism}}

An aspect that I will not dwell on is the total interconnectedness and determinism of the universe. The experiment you plan to do is not arbitrary, but is built into the initial conditions, initial conditions not only within the range of your personal perception but of the universe as a whole. People with different philosophical preferences may view this as extremely negative or extremely positive. I personally am in the latter camp, but Nature does not always respect preferences.

\subsection{Requiring special states\label{srequire}}

The last topic in this section is why there should be the particular restriction on initial states that selects the ``special'' ones. A partial answer is that if at some point in the distant future there is no substantial grotesqueness, then that will impose the special state restriction for all times. For suppose that there is such a future condition. I claim that the \textit{least unlikely} way for it to happen is to have no grotesqueness for all prior times. This is because once you allow a level of macroscopic superposition it's extremely difficult to undo. The living Schr\"odinger cat may become an experimental animal sent to the moon, and the dead one buried outside the perpetrator's lab (the ``worlds'' have split). Getting them back together coherently is not an option. For this reason, satisfying the non-grotesqueness condition for \textit{all} times is less unlikely. And the only way to do this by unitary evolution is by means of special states (which is essentially the definition of special state). I should remark that although for many individual kinds measurements I've shown that special states exist, it is a serious question whether there are enough so that every final condition has enough richness to be the special initial condition for the next thing that's going to happen. (My personal expectation is that the identity of particles---e.g., all electrons are the same electron---makes satisfying this condition less formidable, but that question is one that I did not pursue quantitatively, pending experimental testing of the ideas.)

I call the boundary condition just discussed a ``partial'' explanation because it only leads to another question: why this future boundary condition? Here my response is speculative and may well reflect limitations of my own imagination as well as contemporary scientific ignorance (cf.\ Boltzmann's explanation the arrow of time \cite{boltzmann} as a fluctuation in an enormously long-lived universe). In the usual many world discussions the image is of steady branching to more and more ``worlds.'' With this picture it is not absurd (but also not necessarily implied) that long ago there were fewer such worlds, perhaps at some early stage just one initial wave function that had no macroscopically different superpositions. (Some would call this a quantum arrow of time.) Now let's imagine a cosmology in which there is an eventual contraction. This does not seem a likely scenario in view of the discovery of accelerated expansion, but in the many speculations on the implications of that discovery, contraction, or even a big crunch, is far from having been ruled out. Under these circumstances it is plausible to argue that the arrow of time is a consequence of space-time geometry, so that the end and the beginning should have roughly the same state, which would be non-grotesque. This is admittedly a lot to swallow. But I would refer to the many revolutions that cosmology has undergone, even since the 1930's discovery of expansion \cite{note:landau}. Or, this condition on states may obtain for reasons that I am totally unable to imagine, just as limited knowledge of cosmology in Boltzmann's day made some of his views on the arrow of time untenable.

\section{Experiment\label{sexperiment}}

\subsection{Properties of the kick\label{skickproperties}}

In Sec.\ \ref{sprobabilities} we looked at a two level spin system passing through a Stern-Gerlach (SG) apparatus. Our purpose was to establish minimal requirements for \textit{any} kind of special state. Now however, we really want to consider the true physical system, the Stern-Gerlach experiment.

As emphasized, our usual perspective is \textit{not} to focus on the dynamics of this system alone, but rather on that of the entire environment necessary for a full description. The richness of the environment is what supports the existence of special states. However, in the present section and in the analysis of Sec.\ \ref{sprobabilities}, a different viewpoint is taken, closer to the way most quantum calculations are done. The environment is in whatever special state it is in, but because this state may be rare, its action on the particle or spin of interest will also be unusual. We focus on that action alone and treat the environment's rare action through an effective Hamiltonian. This Hamiltonian provides the time evolution of the wave function through left multiplication by $\exp\left(-iH_\mathrm{eff}t/\hbar\right)$\@.

The system is prepared by passing a beam of atoms through another Stern-Gerlach apparatus and only that part of the beam having a particular value of angular momentum, say $+\hbar/2$ along a particular direction, is selected and sent on to the next SG apparatus. The second SG apparatus is not (necessarily) oriented in the same direction. Let the direction of motion (aside from the eventual deflection) be in the positive $y$ direction and the gradient of the second SG apparatus be in the $z$ direction. Let $\i,\j$, and $\k$ be unit vectors along the $x,y$, and $z$ axes, respectively. We assume that when entering the second apparatus---which is the one on which we focus---the atom's spin is along the direction ${\bm n} = \k \cos\theta - \j\sin\theta$, for some angle $\theta$\@. As in \Eqref{einitialsterngerlach}, the initial wave function of an atom, when exiting the first SG apparatus, can be taken to be
\be
u_\theta = e^{i\theta\sigma_x/2} \left(\begin{array}{c}1\\0\end{array}\right)
         =  \left(\begin{array}{c}\cos\frac\theta2\\  \noalign{\vskip 1pt}
               i\sin\frac\theta2\end{array}\right)
\label{einitial}
\,,
\ee
consistent with the preparation just specified. It is possible to multiply $u_\theta$ by an arbitrary overall phase or to use density matrices, but this does not affect our conclusions. For the SG experiment the final state should have $|u_\mathrm{f}(1)|=1$ (``\up'') or $|u_\mathrm{f}(2)|=1$ (``\down''), which requires that the angle in \Eqref{einitial} be rotated to become an integer multiple of $\pi$\@. Thus the overall action of the effective Hamiltonian is to add an angle $\phi$ to $\theta/2$ so as to accomplish this goal \cite{note:phivspsi}. We refer to this action of the effective Hamiltonian as a ``kick.'' The kick is thus a left multiplication of the wave function by
\be
e^{-iH_\mathrm{eff}t/\hbar}=e^{i\phi\sigma_x}
\label{ekickaction}
\,,
\ee
bringing it to \up\ or \down. As indicated, the effective Hamiltonian in \Eqref{ekickaction} represents the effect of uncontrollable elements of the environment.

As discussed in Sec.\ \ref{sprobabilities} (and proved in Sec.\ 9.1 of Ref.\ \cite{timebook} or \cite{def}, Sec.\ 4.1) recovery of the Born probabilities requires that the kicks be Cauchy distributed, namely that the probability density for a kick of size $\phi$ should be
\be
C_a(\phi)=\frac{a/\pi}{a^2+\phi^2}
\label{ecauchy}
\,,
\ee
with $a$ a parameter that is small. Moreover, it is a property of this distribution that the least unlikely way to achieve large (compared to $a$) total rotation of the spin is through a single kick.

For \up\ we thus require $\phi=n\pi-\theta/2$ and for \down\ $\phi=(n+1/2)\pi-\theta/2$, with $n=0,\pm1,\pm2,\dots$. Define \be
F_a(\psi)\equiv\sum_{n=0,\pm1,\pm2,\dots}\frac{a/\pi}{a^2+(n\pi-\psi)^2}
\label{eFdef}
\,.
\ee
Then the probability for the two outcomes is
\be
\Pr(\up)=\frac1Z F\left( \frac\theta2\right)
\;, \qquad
\Pr(\down)=\frac1Z F\left(\frac{\theta-\pi}2\right)
\,,
\label{ekickprob}
\ee
with $Z$, the sum of $F$ at the two values, providing normalization. For small $a$ this recovers the standard probabilities. The sums can be done explicitly, but we hold off, since there will be related sums to evaluate and we will do all of them at once.

In searching for evidence of special states, presumably the larger the kick the larger the signal. With this in mind, we calculate the expectation of kick size both conditioned on an outcome and unconditioned. We thus want
\bea
\left\langle\phi\right\rangle_{_{\up}}&=&
                   \frac a{Z\pi} \sum_{n=0,\pm1,\pm2,\dots}\frac{ (n\pi-\frac12\theta)}{a^2+(n\pi-\frac\theta2)^2}
                   \label{eupsum}\\
\left\langle\phi\right\rangle_{_{\down}}&=&
                   \frac a{Z\pi} \sum_{n=0,\pm1,\pm2,\dots}
                   \frac{ (n\pi-\frac{\theta-\pi}2)}{a^2+(n\pi-\frac{\theta-\pi}2)^2}
\label{edownsum}
\eea
and their sum.

To evaluate Eqs.\ \parenref{eFdef}, \parenref{eupsum} and \parenref{edownsum} consider the following identity \cite{hille}
\be
\frac1{\tan z}=\sum_{n=-\infty}^\infty \frac1{z-n\pi}
\,,
\label{etanidentity}
\ee
where $n$ runs over the integers. The poles of one over the tangent function occur at multiples of $\pi$ and the residues are unity. Let $z=\theta+ia$\@. Using elementary relations we write the real and imaginary parts of \Eqref{etanidentity},
\be
\frac{\tan\theta}{\tan^2\theta\cosh^2a+\sinh^2 a}
   =\sum_n\frac{\theta-n\pi}{(\theta-n\pi)^2+a^2}
\label{erealpart}
\,,\ee
\be
\frac{\tanh a}{\tanh^2 a\cos^2\theta+\sin^2\theta}
=   \sum_n    \frac{a}{(\theta-n\pi)^2+a^2}
\label{eimagpart}
\,.\ee
From \Eqref{eimagpart} we get the following information: $Z=\frac{4a}\pi\frac1{\sin^2\theta}$ and for sufficiently small $a$, $\Pr(\down)/\Pr(\up)=\tan^2(\theta/2)$, as it should.

\refstepcounter{remark} \label{rfinitea}
\smallskip\noindent\textsf{Remark \arabic{remark}}:~
As mentioned in Sec.\ \ref{sprobabilities} and explicitly calculated in \cite{def, timebook}, for $a$ \textit{not} negligible there will be a deviation from standard probabilities. This imposes a restriction on $a$, but does not provide an experimental test since, in the absence of physical specifics, there is no information on the size of~$a$\@.

\Eqref{erealpart} gives the sums used in the expectations of the kick-angles and yields
\bea
\left\langle\phi\right\rangle_{_{\up}}&=&
                   -\sin\frac\theta2\cos^3\frac\theta2
                   \label{eupsumeval}\\
\left\langle\phi\right\rangle_{_{\down}}&=&
                   -\sin^3\frac\theta2\cos\frac\theta2
\label{edownsumeval}
\eea
If $\theta\approx0$ there is no specializing, so the expected kick size for those measured as \up\ goes to zero. Surprisingly perhaps for those measured as \down\ the expectation is even smaller. This is because although the kicks (however few) are larger, they are as likely to be positive as negative.

According to Eqs.\ \parenref{eupsumeval} and \parenref{edownsumeval} the average kick size is order unity, although given the quirks of the L\'evy distributions, this was not a foregone conclusion. Looking at Eqs.\ (\ref{eupsum}) and  (\ref{edownsum}) it is clear that moments higher than the first do not exist (the first moment is borderline), so that it is conceivable that with experimental studies that focus on large kicks other information may be gleaned.

\refstepcounter{remark} \label{rconvergence}
\smallskip\noindent\textsf{Remark \arabic{remark}}:~
Three of the series that we have considered, Eqs.\ \parenref{eupsum}, \parenref{edownsum} and  \parenref{etanidentity}, are only conditionally convergent. As Hille \cite{hille} remarks in connection with \Eqref{etanidentity}, one can add $1/n$ ($n\ne0$) to each summand to obtain absolute convergence, or what is essentially the same thing, choose to combine positive and negative $n$ terms before summing the infinite series.

\refstepcounter{remark} \label{rquirks}
\smallskip\noindent\textsf{Remark \arabic{remark}}:~
If one performs a series of experiments and manages to measure the kick in each of them, the average will not converge to the results of Eqs. \parenref{eupsumeval} or \parenref{edownsumeval}. This is where the ``quirks'' of the L\'evy distribution enter. As remarked, some of our series are not absolutely convergent and the distribution is not self-averaging. In fact the average of many measurements has the same probability distribution as a single measurement. This can be useful for the experimentalist looking for the effect, since even with averaging there is no suppression of large magnitude kicks. The use of the average might be thought of as the setting of the scale but in fact the only scale is $a$, which is taken to be small. By conditioning on large events, $a$ disappears and there is really no scale.

Depending on experimental setup, it is possible to optimize the angle for maximum signal. For example, suppose one is able to sort particles according to outcome. Then to optimize as a function of $\theta$, one would consider the strength of the field needed for (say) \up, times the probability of \up\@.  This is proportional to $F(\theta) \equiv\cos^2(\theta/2) \langle\phi\rangle_{_{\up}}$. The derivative of this function, $F'=(1/2)\cos^4(\theta/2)\left(\cos^2(\theta/2) - 5\sin^2(\theta/2)\right)$, vanishes for $\theta=\pi$ and $\theta=2\tan^{-1}(1/\sqrt5)\approx48^\circ$\@. Both are stationary points, but the maximum is the second value, $48^\circ$\@. On the other hand, one may send in a large number of particles and simply want to maximize the (absolute value of) the total, $|\cos^2(\theta/2) \langle\phi\rangle_{_{\up}} +\sin^2(\theta/2) \langle\phi\rangle_{_{\down}}|$\@. This gives $F\equiv (1/2)(\sin\theta)[1-(1/2)\sin^2\theta]$ which has a shallow minimum at $90^\circ$ and maxima symmetric about this minimum, one of them being at $\theta=\sin^{-1}(\sqrt{2/3})\approx 55^\circ$\@.

It follows that there is not a lot of profit in fine tuning the optimization. However, what is more significant is that there is a definite $\theta$ dependence. Thus if $\theta$ is varied between 0 and $\pi/2$ one could compare a $\theta\approx0$ no-signal situation (no special state is needed) with a positive signal situation, say at $\theta\approx50^0$\@.

\subsubsection{Strength of the field inducing the kick\label{sfieldstrength}}

For a spin about to enter a Stern-Gerlach apparatus, the effective part of $H_\mathrm{eff}$ of \Eqref{ekickaction} involves a magnetic field, $\bm B$\@. For the kick angle $\phi$ to have characteristic size unity we require
\be
\left|\frac{H_{\mathrm{eff}}\Delta t}{\hbar}\right|
= \left|\frac{{\bm \mu}\cdot {\bm B} \Delta t}\hbar\right|=\frac12 \left|\phi\,\right|\sim1
\,,
\label{ebstrength}
\ee
where $\Delta t$ is the duration of the field's interaction with the spin. The quantity $\bm\mu$ is essentially the electron magnetic moment; taking its magnitude to be the Bohr magneton, implies
\be
B\Delta t \sim 10^{-11} \,\hbox{Ts}
\,.
\label{efieldstrength}
\ee
To evaluate $B$ requires an estimate of $\Delta t$, in turn requiring some picture of the nature of the interaction. At this stage, two possibilities present themselves. The field may be connected to the strong magnetic field the atom experiences in approaching and passing through the magnets. Or the field could be something separate, carried perhaps by an externally arriving photon.

We first consider a possible association with the SG field. A conservative estimate would be interaction durations of a few ms, in which case field strengths would be about $10^{-8}\,$T, which is well within the range of macroscopic measurement. However, this is probably too conservative. In a typical SG experiment the Ag or K atoms are moving at about 1$\,$km/s. If the kick takes place within about 10\littlespace cm, then $\Delta t\sim1\,\mu$s and the field strength would be on the order of 0.1\littlespace G, something your compass needle could discern. \textit{Note added after publication: This estimate is in error. See}~\cite{note:erratum}.

As far as an electric field generated by this transient field, Maxwell's equations suggest $E\sim \frac {L B}{T}$, where $L$ is the characteristic scale for the spatial variation of $E$ and $T$ the time scale for variation of $B$. If $L\sim10^{-1}\,$m and $T\sim1\,\mu$s, we find an electric field on the order of $1\,$V$\!$/m, also easily measurable. Another estimate in this connection uses $L/\Delta t\sim v=1\,$km/s. Thus $E\sim\frac{LB}{\Delta t}\sim \frac{LB\Delta t}{(\Delta t)^2} = \frac{vB\Delta t}{L} = \frac{10^{10}10^{-11}}{L} \sim\frac{10^{-1}}{L}$.

\smallskip

Now consider an outside photon, not necessarily related to the magnetic fields of the SG apparatus. An estimate of this photon's energy can be made in terms of the time of interaction: since $\bm \mu\cdot\bm B$ is an energy, by \Eqref{ebstrength} that energy should be roughly $\hbar/\Delta t$\@. If $\Delta t$ is a characteristic electromagnetic interaction time, $10^{-16}\,$s, this gives an energy on the order of 5\littlespace eV\@.

\subsubsection{Magnetic fields along the particle path\label{sfields}}

A convenient way to study the field in the Stern-Gerlach apparatus \cite{MIT} is to replace the magnets (for purposes of calculation) by a pair of infinite parallel wires with currents flowing in opposite directions. The magnitude of the field is then constant on (circular) cylindrical surfaces for distances large in comparison to the wire separation. This matches the field seen by the passing particle if the pole pieces have the shape of those cylinders. As desired, this magnetic field has a steep gradient perpendicular to the cylindrical surfaces.

\begin{figure}
\includegraphics[height=.3\textheight]{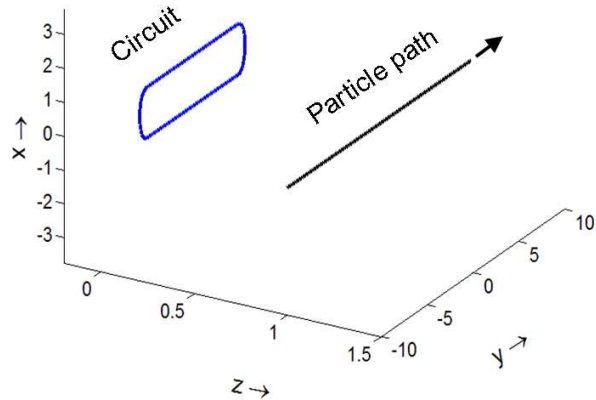}
\caption{Geometrical configuration. The separation of the wires is $s$\@. The particle moves in the positive $y$ direction in the plane $x=0$ and at a positive, essentially constant $z$ value that is larger than $s$\@. Looking at the circuit from positive $z$, \Eqref{estraightwire} corresponds to a current moving in the clockwise direction.}
\label{fwires}\end{figure}

Our interest is not so much in the field within the magnet as the field seen by the atom as it approaches the magnet, moving in the positive $y$ direction. This will certainly depend on the specifics of the magnet, but to get a handle on those fields and to go beyond dimensional analysis, we study the \textit{finite length} magnetic field by simulating the actual field by one generated by a current loop that consists of two wires, but now they are finite. They extend for the length of the magnet and are joined at each end by a semicircular loop (completing the circuit). Fig.\ \ref{fwires} illustrates the following geometry: The circuit is in the $x$-$y$ plane ($z=0$). The straight-wire portions run from $y=-L/2$ to $+L/2$, the upper portion at $x=+s$, the lower one at $x=-s$\@. The semicircles at each end (also in the $x$-$y$ plane with $z=0$) are of radius $s$\@. The particle trajectory is in the direction of increasing $y$ and parallel to the $y$-axis. It has $x=0$ and a value of $z$ large enough so that in its neighborhood the contour lines of the field are essentially circles in the $x$-$z$ plane. The field at a point $\R=y\j+z\k$ is given by the following integral:
\be
{\bm B}(\R)=\frac{\mu_0I}{4\pi}\oint_\Gamma d{\bm r} \times\frac{(\r-\R)}{~|\r-\R|^{3/2}}
\ee
where $I$ is the current and SI units are used. Now the particle \textit{is} deflected in the positive or negative $z$ direction (that's the point of the experiment). But there will also be some spread of the beam in the $x$ direction whose consequences for the field we will evaluate to lowest order. The contour, $\Gamma$, consists of four parts, the top (``T'') portion of the wire parallel to the $y$ axis, the bottom (''B'') portion, the right semicircle (``R,'' $y=L/2$) and the left semicircle (``L,'' $y=-L/2$). For $\R=y\j+z\k$ (which is the plane $x=0$), the straight wire portions can be fully integrated and give
\bea
{\bm B}_{\mathrm{T}\,\&\,\mathrm{B}}(\R)&=& \frac{\mu_0}{4\pi}
    I\int_{-L/2}^{L/2} d\eta\, \j \times\frac{s\i +(\eta-y) \j-z\k}{\left[s^2+z^2+(\eta-y)^2\right]^{3/2}}
    \nonumber \\
      \noalign{\smallskip}
        &&\qquad\qquad
    +\left\{I\to-I\;\&\;s\to-s\right\} \nonumber \\
    \noalign{\smallskip}
    &=&\frac{\mu_0}{4\pi}\left(\frac{-2Is\k}{s^2+z^2}\right)\left[\sin\theta_2-\sin\theta_1\right]
\,,
\label{estraightwire}
\eea
where $\tan\theta_{\left(2\atop1\right)}=(-y\pm L/2)/\sqrt{z^2+s^2}$\@. We also present the first order correction for small $x$, i.e., the observation point $\R$ becomes $x\i+y\j+z\k$\@. The additional term is of the form $x\,\partial {\bm B}/\partial x|_{x=0}$\@. After a bit of calculation one obtains
\be
\left.\frac{\partial{\bm B}_{\mathrm{T}\,\&\,\mathrm{B}}(\R)}{\partial x}\right|_{x=0}=
       -\frac{\mu_0I}{4\pi}2sz\i\left[A_+-A_-\right]
  \,,
\label{estraightwirex}
\ee
where $A=\lambda\frac{(2\lambda^2+3b^2)}{b^4(\lambda^2+b^2)^{3/2}}$ ($\pm$ implicit on $A$ and $\lambda$), with $\lambda_{\pm}=\pm\frac L2-y$ and $b^2=s^2+z^2$. Because of the $z$ dependence, vertical ($x$) spread in the beam will cause (unwanted) blurring of the spin-induced splitting.

The field from the two semicircular portions does not have a general closed form solution but analytic information can still be obtained. The length of the path within the magnet, $L$, will be assumed long enough so that we need consider only one semicircle at a time. Moreover, with respect to the SG apparatus on which we focus (the second) the field on exit is irrelevant, since at that stage only location is measured, not spin. Nevertheless, the exit field will play a role for the first apparatus, because it can change what we assume is the incoming state. Qualitatively though, the possible effects will be the same.

A point on the semicircular portion of the wire near $-L/2$ is given by $\r=-\frac L2\j\allowbreak-s\left(\j\sin\psi+\i\cos\psi\right)$, with $\psi$ running from 0 to $\pi$\@. For clockwise circulating current (as viewed from positive $z$) $d\psi$ is in the direction of the current. After a bit of calculation we obtain an expression for the left semicircular (``L'') contribution
\be
\B_{\mathrm{L}}(\R)=\frac{\mu_0 I}{4\pi}\int_0^\pi s\,d\psi\,
 \frac{z\left(\i\cos\psi+\j\sin\psi \right)-\k\left(\bar y\sin\psi+x\cos\psi +s\right)}
              {\left[ x^2+\bar y^2 +s^2 +2s\left(\bar y\sin\psi+x\cos\psi\right) \right]^{3/2}}
\label{ebleftsemicircular}
\,,
\ee
where $\bar y\equiv y+\frac L2$\@. For purposes of studying the effective Hamiltonian, \Eqref{ekickaction}, we are only interested in the $x$-component of this field. Specializing to $x=0$, the integral can be performed, yielding
\be
B_{\mathrm{L}\,x}(\R)=\frac{\mu_0 I}{\pi}\left[\frac1{\sqrt{\bar y^2+s^2+z^2}} -  \frac1{\sqrt{ (\bar y+s)^2+z^2}} \right]
\,.
\label{ebxleftsemicircular}
\ee
As the atom approaches the magnet, this field rotates the spin one way and then the other. The magnitude of this field is substantial. Rewrite the field as
\be
B_{\mathrm{L}\,x}(\R)=\frac{\mu_0 I }{4\pi}\frac1z\left[\frac4{\sqrt{\left(\frac{\bar y}z\right)^2+\left(\frac s z\right)^2+1}} -  \frac4{\sqrt{ \left(\frac{\bar y+s}z\right)^2+1}} \right]
\,.
\label{ebxleftsemicircular2}
\ee
The dimensionless quantity in the square brackets has a maximum of about $1/2$ for $s/z \sim 0.75$, which is approximately the value in the experiment of Ref.\ \cite{MIT}. Comparing \Eqref{estraightwire} and \Eqref{ebxleftsemicircular2} it is seen that the external field reaches almost half the field value inside the magnets.

\subsection{Detection scenarios\label{sdetection}}

The general strategy is to send in atoms with spins at (say) 50$^\circ$ relative to the $z$-axis (tilted along the $y$-axis) and to send them in at 0$^\circ$ \cite{note:tilty}. Comparison of the two cases should show additional ``random'' activity---noise---when they are at the non-zero angle. At 0$^\circ$ no kicks are necessary to drive the spins into a single beam for the SG experiment. At 50$^\circ$ they will all need to be sent one way or the other. The actual rotating of the spins would not itself be visible, but related and additional fields should be present. The idea is that there should be ``collateral damage,'' by which is meant that the photon or field fluctuation is not perfectly matched to accomplish its rotational task and nothing more. As discussed at length in Ref.\ \cite{timebook}, in generating a special state one seeks the \textit{least unlikely} of them. A fundamental assumption in the present proposal is that a perfect match is less likely than an imperfect one. In addition, by virtue of Maxwell's equations, there are compulsory electric fields alongside the magnetic fields that rotate the spin.

Ways to fine-tune the strategy above may certainly exist. For example, if the signal of a kick can be correlated with a particular atom (which goes either \up\ or \down), differences in signal rates for different angles can be further exploited.

\subsubsection{Scenario when the fields are generated by the SG magnets\label{ssgdetection}}

One issue is the stability of the fields. The fields needed for rotating the spins are on the order of 1$\,$G, while the magnet is maintaining a field of roughly 5000\littlespace G\@. One thus needs field measurements with better than 0.1\% accuracy. It should also be recalled that the preparation of the spin at some particular angle is accomplished by means of a earlier SG setup. Kicks can occur in the first as well as the second magnet. The rotating fields for the magnets (meaning, for the example studied, fields in the $x$-direction) are also different for different atoms because of finite beam width (cf.\ \Eqref{ebxleftsemicircular2} where there is $z$-dependence in the field).

Furthermore, the magnetic fields that can rotate the spin are necessarily accompanied by electric fields
since the variety of rotation directions through the magnet (for $\theta\ne0$) demands time-dependent variation of~$\bm B$\@. With an atomic velocity of 1\littlespace km/s, a conservative estimate puts these fields on the order of 1 or more~V/m.

For an atomic beam, there may be additional effects. Many atoms pass through the magnet at roughly the same time. Not all of them are rotated the same way, so that rapid variation of the magnetic field would be required (along with the electric fields just discussed). In addition, the ``least unlikely'' principle suggests that there would be a tendency for bunching in the output, that is there would be short-time correlations in \up\ or \down~outcomes. The rationale is that a single large fluctuation is more likely than two independent ones.

\subsubsection{Scenario when the fields are generated by external photons\label{sphotondetection}}

Our rough estimate for photon energy was in the eV range, visible or UV light when the kick drives the spin around many times (as is occasionally expected, given the Cauchy distribution). Individual photons in this energy range should be easy to detect.

It should be pointed out though that the estimates of Sec.~\ref{sfieldstrength} are only that---estimates. A general scale is established. However, the properties of the Cauchy distribution imply that this scale will often be vastly exceeded. For this reason I do not go beyond the semiclassical assumption, implicit in that calculation, that the field acts on the atom, but not vice versa. For atom-photon scattering one should in principle work in a QED context. My assumption is that both incoming photon and outgoing photon will all be on the scale of the estimate.

\section{Discussion\label{sdiscussion}}

There are three issues to be taken up in this discussion: 1)~~Comments on the plausibility of the overall theory. 2)~~Review of the nature and assumptions in the experimental test. 3)~~The possibility of other tests.

Concerning the special state theory, I think that Bohr's criterion of being ``crazy enough'' is satisfied \cite{note:bohr}. Personally I have no problem with the restriction on initial states, nor on the idea of what is sometimes called a ``block universe,'' one in which past and future are all part of a unified space-time (and maybe more) history. Where my credibility is stretched is the possibility that there are \textit{so many} microstates that specialness is possible again and again and again. On the other hand, I am sufficiently unhappy with other quantum measurement ideas, either giving up unitarity or having many worlds or giving up the idea that the wave function is any more than a computational tool, that I am prepared to entertain this ``crazy enough'' idea.

The proposed experiment would involve two sets of Stern-Gerlach apparatus, one for preparation, one for measurement. The calculations in this article leave open two possibilities for the detection of a signal accompanying the rotation of the atom's spin. In one case, there would need to be high quality light sensors along the path between them (which the experimentalist must therefore maintain in darkness). In the other, precise measurements of the magnetic field (as well as stability of that field) would be necessary. Alternatively electric fields could be measured close to the entry to the magnets. It is also possible that bunching effects would be detected in measurements of atom positions.

Our proposals are based on a number of assumptions. For photon measurements, we expect the energy of the emitted photon to be in the eV range. This is based on no more than the fact that the usual time scale for electromagnetic interactions is $10^{-16}\,$s. I can easily imagine an order of magnitude correction in either direction. However, the range of ``kick'' sizes is also great, so that even if, say, the bulk of the photons landed in the infrared, some would be visible. Moreover, there are sensors for these other energy ranges. Another assumption is the concept of what I have called ``collateral damage.'' Namely, if the spin is to be rotated by a specific amount, it is likely that the field or photon doing the job is not exactly tailored to do only that, but rather would have some other energy value and would carry away the excess. Moreover, since the strength of the needed kick has a long-tail distribution, the excesses, presumably on the same scale, would have the same distribution. In addition, if the rotating field is that of the magnet, even if there is little or no excess in the magnetization field, a significant electric field (demanded by Maxwell's equations) would still appear. The nature of the demand for the electric field implies that it too be Cauchy distributed. There are of course other assumptions, such as identifying the location of the kick as the atom's path before being well into the second SG magnet, but they seem to me more secure hypotheses.

One might also ask, does the measurement of the ``kick'' on the path of the particle already fix the outcome, in the same way that checking which slit a particle goes through can destroy the interference pattern in a two-slit experiment. Analyzing this question requires determining whether the upstream (i.e., before entering the second SG magnet) measurement can actually predict the outcome, which in turn requires a more quantitative estimate of the expected signal. However, from the standpoint of confirming the theory described above, there are two aspects of the suggested tests that are significant even if predictive information could be deduced from the upstream measurement. First, the contrast between 0$^\circ$ and 50$^\circ$ entry beams (the angles are the orientation of the atoms relative to the $z$ axis of the second SG magnet) would exist whether or not the spin localization (``space quantization'') were observed. At 0$^\circ$ there would be \textit{no} signal, not photons, not electric fields; at 50$^\circ$ there would be such a signal whether or not the usual SG splitting were observed. Secondly the observation of a Cauchy distribution in the noise would also be support for this theory, since nothing in the Copenhagen interpretation involves long tailed distributions.

The last issue concerns other possible two- (or more) state observations. The beam splitters and polarizers, used for example in \cite{jacques} and working with photons instead of atoms, do jobs similar to that of the Stern-Gerlach apparatus and may be simpler to set up. I have not analyzed such experiments because I have less confidence in being able to identify where the least unlikely changes in the photon would take place. Partly this is my own ignorance and partly this reflects the greater complexity in, say, rotating polarization, involving as it does a medium. But in principle the Cauchy distributed noise should appear whenever a selection of macroscopic states is demanded. If this can conveniently be matched with cases where no selection is needed (as in sending in beams in the SG experiment oriented at 0$^\circ$ and 50$^\circ$) then the comparison should show the differences the special state theory predicts.

\begin{acknowledgments}
The list of those to whom I am grateful for participation in the formation of these ideas is long and can be found in~\cite{timebook}. In developing the new material in this article I have received helpful advice from Bernard Gaveau, Karel Polak, Carlo Rizzo, Marco Roncadelli and Dipanker Roy. I must, however, add to these thanks that any errors, especially when experimental details are under discussion, are my own. I am also grateful for the hospitality of the Max Planck Institute for the Physics of Complex Systems in Dresden, where some of the work on this article took place.
\end{acknowledgments}

\appendix
\section{Search technique for special states\label{smethod}}

I use the notation of the decay narrative in Sec.\ \ref{squantum}. Let the projection operator for the $n$-dimensional subspace of the initially excited atoms be called $P$\@. Let the propagator for the full $(N+n)$-dimensional Hamiltonian ($H$) be called $U$, so that $U=\exp(-iHt_0/\hbar)$, where $t_0$ is the particular time at which the state must be non-grotesque. If the initial state, $\psi_0$, is undecayed then it satisfies $P\psi_0=\psi_0$\@. The probability that at time-$t_0$ it is still undecayed is $S(t_0)=\left|\left|PU\psi_0\right\rangle\right|^2$, the ``survival probability.'' This can be rewritten as $S(t_0)=\left\langle\psi_0\left| C^\dagger C\right|\psi_0\right\rangle$, with $C\equiv PUP$\@. The problem of finding states that decay entirely or do not decay at all becomes the problem of finding eigenvectors of $C^\dagger C$ with eigenvalues near 0 or 1\@. In general for large enough systems (thinking beyond the particular decay model of the Hamiltonian \Eqref{edecayhamiltonian}) there will be many eigenvalues quite close to both limits. For the case at hand (and this is related to the straightness of the line in Fig.\ \ref{fmultileveldecay}) almost all the eigenvalues cluster around zero and one \cite{linear}. The latter property holds when the coupling matrices $\phi$ are essentially constant.

The figures shown in Sec.\ \ref{squantum} are based on numerical calculations.

\bigskip

\noindent\textbf{References}
\vskip-30pt

\begin{thebibliography}{28}
\bibitem[{not({\natexlab{c}})}]{note:calculation}
\bibinfo{note}{We also use probability when---even if the initial point is
  known---it is impractical to calculate the later phase space point, for
  example for chaotic dynamics. Taking this into account would lead to a slight
  restatement of the assertions in the text, but the quantum issues, which are
  the point of the discussion, are the same.}
\bibitem[{not({\natexlab{e}})}]{note:landau}
\bibinfo{note}{L. Landau is said to have characterized cosmologists as, ``Often
  in error; never in doubt.''}
\bibitem[{not({\natexlab{g}})}]{note:tilty}
\bibinfo{note}{They could as well tilt along the $x$-axis, although details of
  our calculation would be slightly different. Note too, that for, say
  50$^\circ$, one would select from the output of the first SG setup. However,
  for the 0$^\circ$ experiment this is not necessary, since there would still
  be an absence of ``noise'' even if both orientations are incoherently
  recombined and sent through the second SG setup.}
\end{thebibliography}

\end{document}